# Simultaneous Motion Monitoring and Truth-In-Delivery Analysis

# Imaging Framework for MR-guided Radiotherapy

Nikolai J. Mickevicius[1], Xinfeng Chen[1], Zachary Boyd[3], Hannah J. Lee[4], Geoffrey S. Ibbott[4], Eric S. Paulson[1,2,3]

[1]Radiation Oncology, [2]Radiology, and [3]Biophysics, Medical College of Wisconsin, Milwaukee, Wisconsin, United States
[4]Radiation Physics, MD Anderson Cancer Center, Houston, Texas, United States

Abstract

Intrafraction motion (i.e., motion occurring during a treatment session) can play a pivotal role in the success of abdominal and thoracic radiation therapy. Hybrid magnetic resonance-guided radiotherapy systems have the potential to control for intrafraction motion. Recently, we introduced an MRI sequence capable of acquiring real-time cine imaging in two orthogonal planes (SOPI). We extend SOPI here to permit dynamic updating of slice positions in one plane while keeping the other plane position fixed. In this implementation, cine images from the static plane are used for motion monitoring and as image navigators to sort stepped images in the other plane, producing dynamic 4D image volumes for use in dose reconstruction. A custom 3D-printed target, designed to mimic the pancreas and duodenum and filled with radiochromic FXG gel, was interfaced to the dynamic motion phantom. 4D-SOPI was acquired in a dynamic motion phantom driven by an actual patient respiratory waveform displaying amplitude/frequency variations and drifting and in a healthy volunteer. Unique 4D-MRI epochs were reconstructed from a time series of phantom motion. Dose from a static 4cmx15cm field was calculated on *each* 4D respiratory phase bin and epoch image, scaled by the time spent in each bin, and then rigidly accumulated. The phantom was then positioned on an Elekta MR Linac and irradiated while moving. Following irradiation, actual dose deposited to the FXG gel was determined by applying a $R_1$ versus dose calibration curve to $R_1$ maps of the phantom. The 4D-SOPI cine images produced a respiratory motion navigator that was highly correlated with the actual phantom motion (CC=0.9981). The mean difference between the accumulated and measured dose inside the target was 4.4% of the maximum prescribed dose. These results provide early validation that 4D-SOPI simultaneously enables real-time motion monitoring and truth-in-delivery analysis for integrated MR-guided radiation therapy (MR-gRT) systems.

**Introduction**

Intrafraction motion (i.e., motion occurring during a treatment session) can play a pivotal role in the success of abdominal and thoracic radiation therapy. With the steep dose gradients typically employed to reduce doses to proximal organs at risk (OAR), large and dynamic translations, rotations, and deformations arising from respiration, peristalsis, organ filling, drifting, or bulk motion can compromise target coverage and obscure OAR doses [1]. Although a number of methods to monitor intrafraction motion [2], [3] and reduce its severity [4]–[6] have been introduced, imaging frameworks that support intrafraction motion *management* and "truth-in-delivery" analysis (i.e., dose reconstruction) are still under development.

Hybrid magnetic resonance-guided radiotherapy (MR-gRT) systems [7]–[10] permit non-ionizing, high soft tissue contrast imaging before (pre-beam), during (beam-on), and after (post-beam) radiotherapy delivery. With its high temporal resolution, two-dimensional (2D) cine MRI is a natural choice for continuous, real-time, intrafraction motion monitoring during beam-on. However, even with the use of simultaneous multislice (SMS) excitation, cine MRI is only capable of imaging a limited number of 2D slices, which limits its utility for dose calculation.

Four-dimensional (4D) MRI is capable of constructing volumetric (3D) images correlated with specific structure motion or motion surrogates. A plethora of 4D-MRI methods have been introduced, including prospective and retrospective multi-slice 2D [11]–[15] and retrospective 3D approaches [16]–[18]. Although 4D-MRI can resolve motion in three dimensions, several motion cycles are required to meet the data sufficiency conditions to construct 4D volumes, which limits the utility of 4D-MRI for real-time intrafraction motion monitoring.

Recently, the combination of pre-beam 4D-MRI and beam-on cine imaging has been proposed as a dose reconstruction imaging framework for hybrid MR-gRT systems. In this approach, a motion model generated from the 4D-MRI is dynamically updated throughout the treatment fraction through deformable image registration (DIR) with the cine images[19], [20]. The dynamic update of the motion model facilitates generation of a synthetic 3D volume at the temporal frame rate of the cine images, while also permitting control for deviations from the average respiratory cycle determined from the pre-beam 4D-MRI. While this is a powerful method, an approach that does not rely on the accuracy of DIR may be desired.

If the use of DIR is to be avoided, dynamically updating 4D-MRIs must be reconstructed throughout the treatment fraction. A straightforward approach would be to acquire continuous 3D data (e.g., golden angle radial stack-of-stars [21]) during beam-on and retrospectively reconstruct 4D-MRI epochs that account for short-term variations in respiratory pattern or physiological effects such as organ filling. However, the ability to acquire real-time images for intrafraction motion *management* (e.g., gating, tracking, trailing) is lost with this approach.

Recently, we introduced a method capable of acquiring real-time cine MR images from slices in orthogonal planes simultaneously (SOPI) [22]. By performing sequential excitation and simultaneous phase encoding of orthogonal slices, signals originating from packets of SMS slices within orthogonal planes can be acquired within the same repetition time (TR). Simultaneous image refocusing (SIR) and/or parallel imaging methods can then be used to separate slices between and within orthogonal slice groups during reconstruction.

In this work we extend the SOPI method to enable simultaneous real-time motion monitoring and construction of dynamic, self-navigated 4D-MRI epochs during an MR-gRT treatment fraction. The goals of the present work were three-fold. First, test the ability of 4D-SOPI to acquire respiratory phase-resolved 4D-MRI volumes in a dynamic motion phantom and *in vivo*. Secondly, test the ability of 4D-SOPI to reconstruct serial, self-navigated 4D-MRI epochs throughout a scan duration consistent with the length of an intensity modulated radiotherapy (IMRT) treatment fraction. Finally, test and validate the use of 4D-SOPI for dose reconstruction in a dynamic motion phantom.

**Methods**

*Pulse Sequence*

Two variations of the 4D-SOPI pulse sequence were implemented. The use of an SMS factor of two was used in the imaging slice group throughout this study. The first variant, referred to as one-plane 4D-SOPI (1PL-4D-SOPI), acquires a cine navigator slice in one orthogonal plane while stepping through groups of SMS imaging slices in the second orthogonal slice plane. The two-plane variation (2PL-4D-SOPI) acquires one cine navigator slice in each of the orthogonal planes while acquiring a single imaging slice at a time. These two sequence variants are depicted in Figure 1.

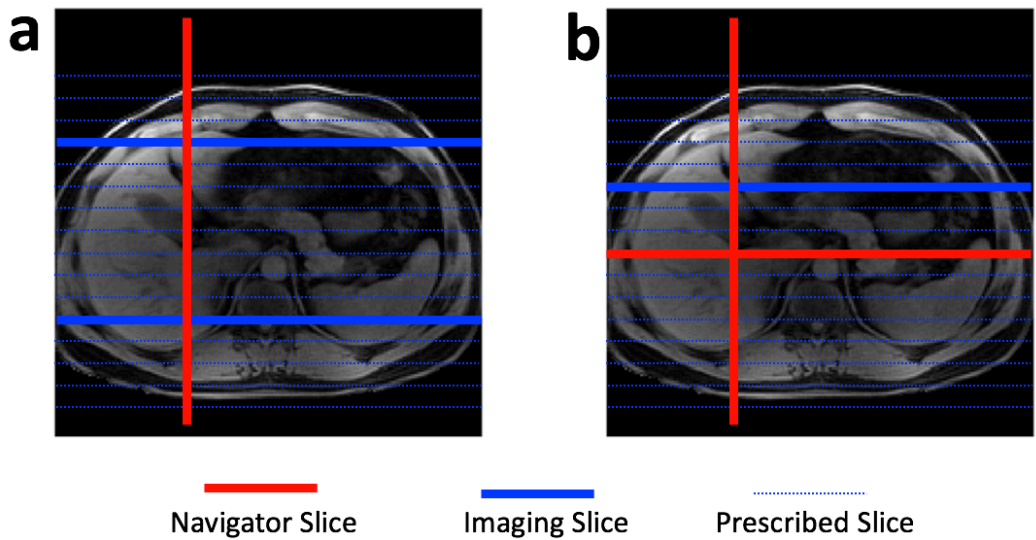

Figure 1. Examples of coronal 4D-SOPI slice prescriptions with SMS=2. (a) One-plane (1PL) navigator 4D-SOPI. Data from a sagittal navigator slice (red) and two simultaneously excited coronal imaging slices (blue) are acquired within each TR. The spacing between the coronal imaging slices is kept constant, but the imaging slice positions are stepped dynamically during acquisition. (b) Two-plane (2PL) navigator 4D-SOPI. Data from one sagittal navigator slice, one coronal navigator slice, and one coronal imaging slice are acquired within each TR. The spacing between the simultaneously excited slices in the coronal plane varies depending on the location of the imaging slice.

The pulse sequence timing diagram for 4D-SOPI is shown in Figure 2. It is based on the previously published nETE SOPI sequence[22]. The slice-select axis of one slice group lies on the phase encoding axis of the other. Both orthogonal slice groups share a frequency encoding direction. In this study, the first RF pulse played out within each TR corresponds to the navigator slice which is orthogonal to the 4D imaging slices. The second RF pulse is a multiband pulse which excites two slices in the 4D imaging plane simultaneously. To spatially encode these slice groups simultaneously, gradient lobe areas must be calculated based on the area of the current phase encoding gradient area, P. The expressions to calculate these areas are as follows:

$$B = -A_2 - P \qquad [1]$$

$$E = -D_2 \qquad [2]$$

$$C = -D_1 + P \qquad [3]$$

$$F = -I \qquad [4]$$

$$G = -H \qquad [5]$$

With the above expressions for the frequency encoding prephaser gradients, F and G, the use of an extended frequency encoding gradient can be used to separate the signals from the orthogonal slice groups according to the SIR principle. The aliased signals from the SMS slices remain entangled within the gradient echo corresponding to the 4D imaging slice packet.

The RF pulses were implemented as phase-tagged sinc pulses with duration 0.6 ms and bandwidth 2 kHz. Phase tags were used to apply CAIPIRINHA phase ramps to achieve a FOV/2 shift between simultaneously excited slices within the 4D imaging plane. Additionally, the SMS slices were excited in quadrature to avoid interference and to facilitate a phase constrained reconstruction. To accelerate the acquisition of each frame, an in-plane reduction R=3 and 5/8 partial Fourier were used.

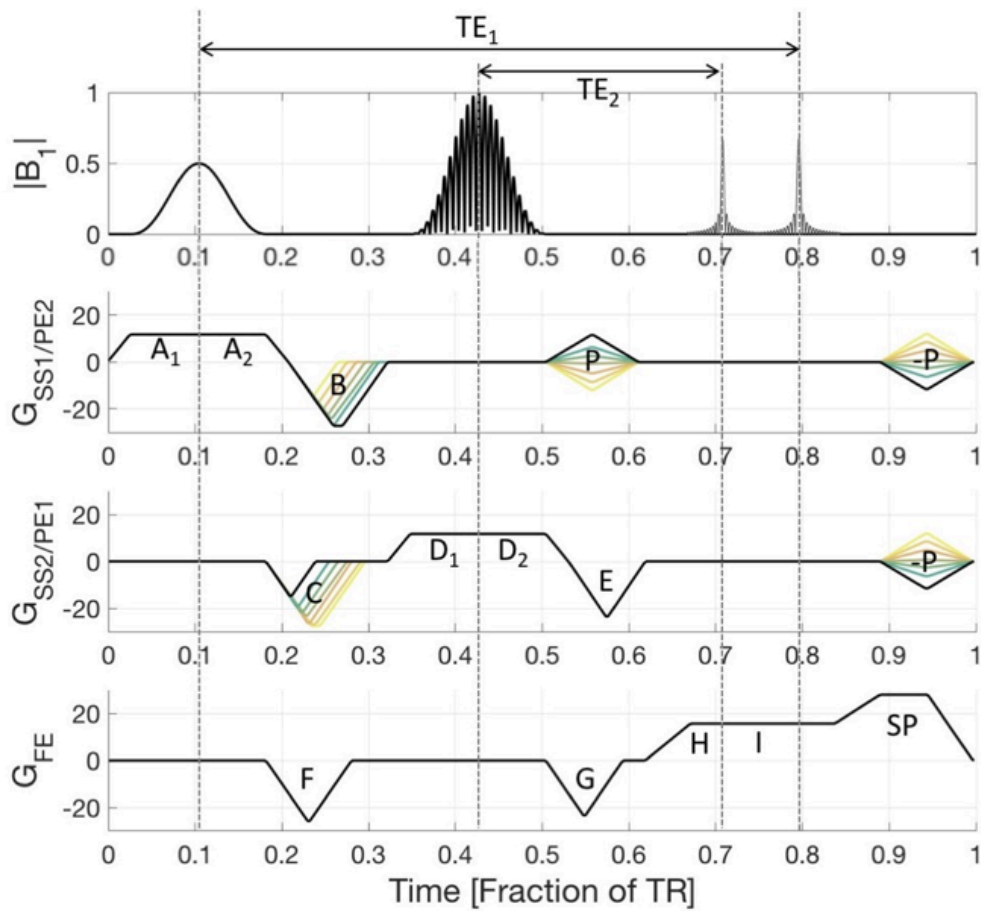

Figure 2. 4D-SOPI pulse sequence timing diagram. The orthogonal slice packets are acquired with non-equal echo times and are separated via a simultaneous image refocusing (SIR) readout. The areas of the B and C gradient lobes are dependent on the current phase encoding moment, P. The phase encoding axis of one orthogonal slice group lies on the slice select axis of the other orthogonal slice group. The two slice groups are frequency encoded on a common axis. A single-band RF pulse is played out to excite a slice in the navigator plane while a multiband RF pulse is used to excite slices in the imaging plane.

*Image Reconstruction*

      To reconstruct the navigator images for the 1PL-4D-SOPI acquisition, a GRAPPA interpolation kernel was calculated from fully sampled autocalibrating signal (ACS) lines near the center of k-space. This kernel was used to interpolate the skipped phase

encoding lines. Subsequently, a homodyne partial Fourier algorithm was used to restore the conjugate symmetric portion of k-space.

A phase constrained 2D-SENSE-GRAPPA algorithm was used to separate signals from simultaneously excited slices in the 4D imaging slice group. This kernel was calculated from a low-resolution calibration scan. The kernel is applied to the undersampled portions of the k-space of an extended readout FOV representation of the SMS slices. The virtual conjugate coil (VCC) method was used to exploit the phase differences between the SMS slices. The same homodyne reconstruction algorithm was used to restore the conjugate symmetric portion of k-space of each SMS slice. The details of the algorithm implementation have been previously described[22].

*Self-Navigated 4D Image Sorting*

An intensity projection through a high-contrast tissue interface was extracted directly from the cine images and plotted over time to create a 1D motion navigator to sort the imaging slices into respiratory phase resolved bins. For each desired 4D-MRI epoch, the motion surrogate was placed into one of four-to-six respiratory bins based on the signal amplitude. The imaging slices acquired concurrently with each frame of the navigator slice were placed into the corresponding respiratory phase. Repeat images within each bin were averaged.

*Phantom Target*

A custom phantom mimicking the pancreas and duodenum was designed and 3D printed with PLA filament (see Figure 3). The pancreas and duodenal structures of the phantom were filled with radiochromic FXG (Fricke xylenol orange) gel [23]. This combination of filament and gel was chosen due to their tissue-equivalent and radiological properties [24]–[29]. An air-filled cavity within the duodenum was incorporated in the phantom design to induce the electron return effect at the duodenal wall during irradiation on a prototype Elekta MR-Linac (Elekta Instruments AB, Stockholm, Sweden) with orthogonal magnetic and irradiation fields. The 3D printed target was interfaced to a dynamic MRI-compatible motion phantom (Model 008M, CIRS, Norfolk, MA).

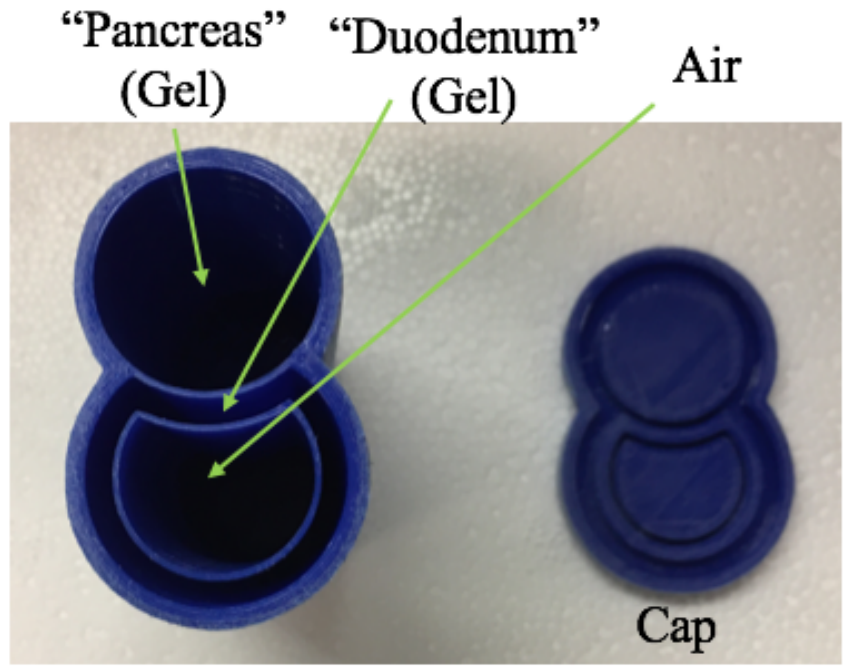

Figure 3. Custom 3D-printed target designed to simulate pancreas and duodenum structures. The pancreas and duodenum cavities were filled with FXG gel. The air-filled cavity was incorporated into the design to induce the electron return effect during irradiation on an MR-Linac with orthogonal magnetic and irradiation fields. The 3D printed target was interfaced to a dynamic MRI-compatible motion phantom (Model 008M, CIRS).

*Phantom Experiment*

The respiratory surrogate waveform extracted from a Varian RPM system during 4D-CT imaging of a pancreas patient is shown in Figure 4. The time duration of the waveform was 7.5 minutes. The waveform was used to drive the dynamic motion phantom in the superior-inferior direction. Dynamic motion phantom images from a 1PL-4D-SOPI FISP scan were acquired on a Siemens 3T Verio scanner (Siemens Healthineers, Erlangen, Germany) with a single sagittal navigator slice and 62 coronal slices with a thickness of 3 mm. The $TE_1$ / $TE_2$ / TR from the scan were 3.5 / 1.5 / 5.3 ms, respectively. A flip angle of 20 degrees was prescribed. The FOV and matrix size were 340 mm and 128x128. A reduction factor of three (R=3) and 5/8 partial Fourier were used for in-plane acceleration. With 48 phase encode lines per frame, the total amount of time to acquire k-space data for an SMS=2 pair of coronal slices and sagittal navigator slice was approximately 0.25 seconds. Three 4D-MRI epochs with five bins were reconstructed, each using a sequential 1/3 of the data. A brief, low-resolution single-band calibration scan was performed to calibrate the 2D-SENSE-GRAPPA interpolation kernels prior to 4D-SOPI acquisition.

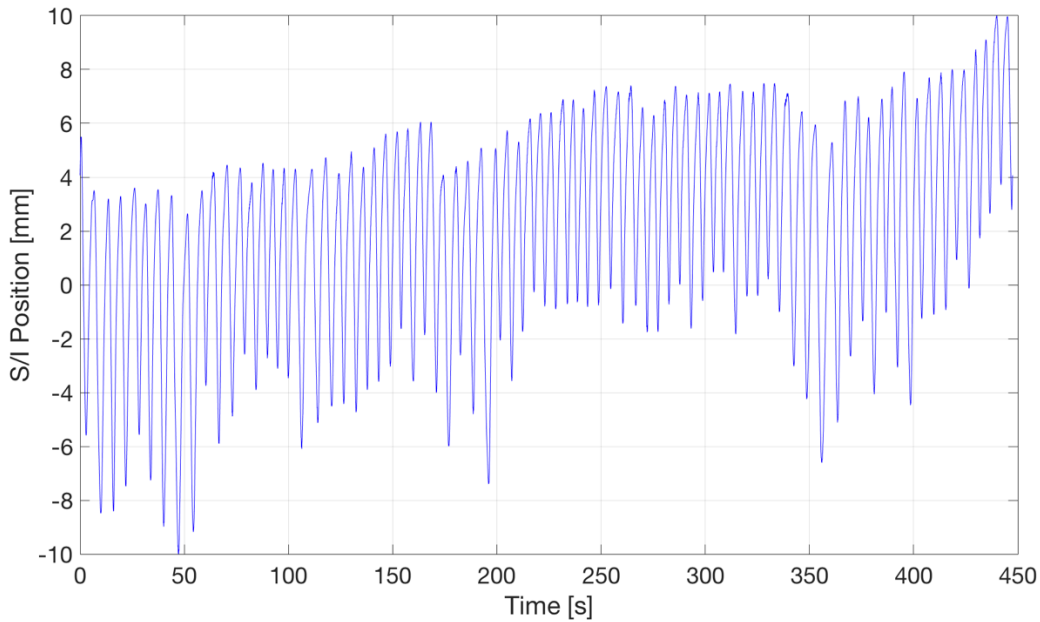

Figure 4. Respiratory waveform obtained during 4D-CT imaging of a pancreas patient and used to drive the dynamic motion phantom during imaging and irradiation. The waveform was chosen because it demonstrates considerable frequency/amplitude variations and drifting.

*Dose Reconstruction Experiment*

An *in vitro* dose reconstruction experiment simulating a hypofractionated pancreas cancer MR-gRT fraction on a prototype Elekta MR-Linac was performed. Fifteen reconstructed 3D volumes of the dynamic motion phantom, obtained from the 1PL-4D-SOPI acquisition described above (5 bins per 4D epoch; 3 total 4D epochs), were resampled to 1mm cubic voxels and then loaded into a research version of the Monaco treatment planning system (TPS) (v5.19.03.03d, Elekta Instruments AB, Stockholm, Sweden). Electron densities of each phantom structure were determined from a prior CT scan of the phantom (Definition AS Open, Siemens Healthineers, Erlangen, Germany) and assigned in Monaco. For each of the fifteen volumes, dose from a single 4cm x 15cm static open field oriented at gantry = 0 degrees was calculated using a 2mm grid spacing and 1% statistical uncertainty per control point. The effects of the MR-Linac cryostat and presence of the transverse 1.5T magnetic field were incorporated during dose calculation. During this process, the position of the beam was kept fixed relative to the static components of the phantom (i.e., the phantom target was at different positions relative to the treatment field according to 4D bin and epoch number). At the nominal MR-Linac output of 432 MU/minute, approximately 3200 MU were required for continuous irradiation over the 7.5-minute duration of the respiratory waveform shown in Figure 4. The relative monitor units (MU) used for dose calculation of each 4D bin and epoch were determined based on the fraction of time the phantom spent in each bin and epoch, shown in Table 1.

Table 1. Fractions of time spent in each 4D-MRI bin and epoch used to determine the fraction of planned 3200 MU to apply to each 4D imaging bin and epoch during dose calculation.

| Fraction of Total MU | Epoch 1 | Epoch 2 | Epoch 3 |
|:---:|:---:|:---:|:---:|
| Bin 1 | 0.0167 | 0.0070 | 0.0113 |
| Bin 2 | 0.0376 | 0.0199 | 0.0355 |
| Bin 3 | 0.0774 | 0.0806 | 0.0538 |
| Bin 4 | 0.1059 | 0.1043 | 0.1898 |
| Bin 5 | 0.0957 | 0.1215 | 0.0430 |

The Monaco TPS was set to deliver 1 cGy/MU to water at 10cm depth and 143.5 cm source-to-axis distance. However, the absolute calibration of the MR-Linac at the same position in water was measured to be 0.8766 cGy/MU at the time of the experiment. Therefore, the doses for each bin and epoch were scaled by 0.8766 to account for the difference between TPS and actual absolute dose. The scaled doses from all fifteen treatment plans were then sent to MIM (v6.7.6, MIM Software, Cleveland, OH). Local rigid registration was performed to align the targets of each 4D bin and epoch to the reference target position of bin 5, epoch 3. After registration, doses were consecutively summed to construct an accumulated dose.

The dynamic motion phantom was then positioned on the MR-Linac. The phantom motion was driven with the identical motion waveform used for the 1PL-4D-SOPI acquisition (i.e., Fig 4) and simultaneously irradiated using a static 4cm x 15cm field with 3200 MU. Following irradiation, the phantom motion was suspended and quantitative spin-lattice relaxation rate ($R_1$) mapping of the dynamic motion phantom and target was performed using DESPOT1 [30] with five flip angles (3,6,10,20,30 degrees) and TR = 20 msec [31]. The 3D $R_1$ map was converted to absolute dose by applying a calibration curve obtained from a calibration experiment in which ten tubes of FXG gel were exposed to known radiation doses ranging linearly from 0 to 4000 cGy and corresponding R1 maps were estimated. The measured, absolute 3D dose distribution of the phantom was then compared against the accumulated dose distribution reconstructed in MIM.

*In Vivo Experiment*

Identical timing and resolution parameters as those used in the phantom experiment were also used to acquire 1PL-4D-SOPI FISP data in the abdomen of a consenting healthy volunteer on a Siemens 3T Verio scanner. Four 4D-MRI epochs of four bins were reconstructed for this scan. Additionally, a brief 2PL-4D-SOPI scan was acquired, and a single volume was reconstructed. The same calibration scan as in the phantom experiment was performed prior to acquisition of 4D-SOPI.

**Results**

An example sagittal navigator image from the phantom experiment is shown in Figure 5a. An intensity projection was extracted from the wall of the target and was plotted over time. The projection and extracted respiratory motion waveform are shown in Figure 5b. The respiratory motion navigator was highly correlated with the actual phantom motion (correlation coefficient = 0.9981). Partial saturation of the current and previous coronal slice groups is apparent within the navigator image in Figure 5a.

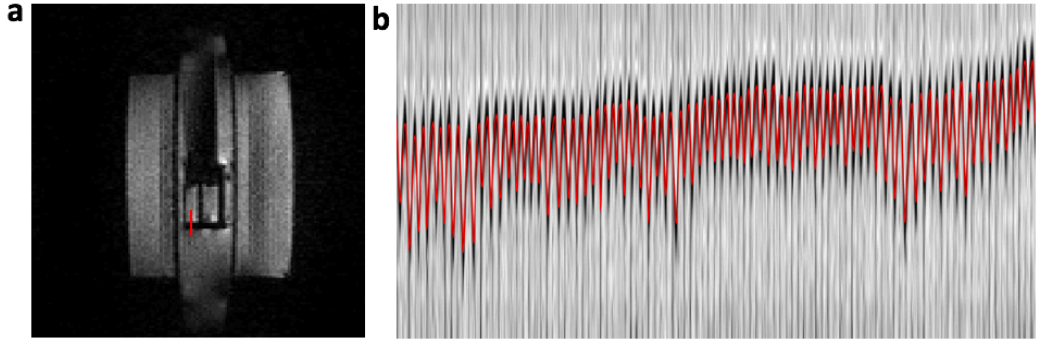

Figure 5. Results from imaging study of dynamic motion phantom driven with patient waveform shown in Fig 3. Example sagittal navigator image (a) and respiratory motion waveform (red time series in b) extracted from an intensity projection (red line in a) placed over the 3D-printed target.

Sagittal reformats of end-inspiratory phase images for each of three 4D-MRI epochs obtained in the phantom experiment are shown in Figure 6. The red reference lines aid in visualizing the differences in target position between the different epochs. As expected, the target position in each of the end-inspiratory epochs differed along the superior / inferior direction due to the drift in the motion waveform used to drive the phantom. Slight slice-mismatch within each epoch is observed.

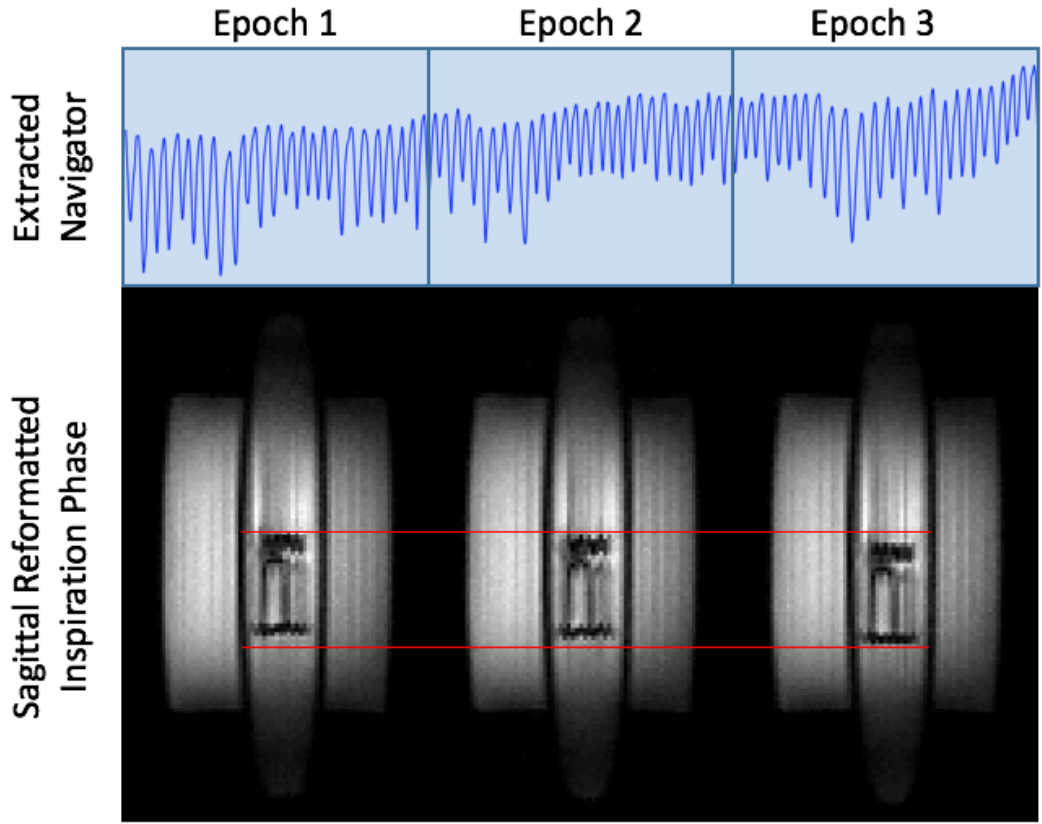

Figure 6. Reconstruction of three separate 4D-MRI epochs (bottom) from extracted navigator (top) obtained during the phantom experiment. Reformatted sagittal images at end-inspiratory phase from the three 4D-MRI epochs are shown in the bottom row. Red reference lines demonstrate that reconstruction of separate 4D-MRI epochs enables drifts to be captured for use in truth-in-delivery analysis.

Figure 7 provides a comparison between a static dose distribution, calculated using the full 3200 MU on a single reference volume, and reconstructed dose distribution. In the reconstructed dose distribution, the dose in the target is smeared along the superior/inferior direction due to the motion of the phantom throughout the experiment.

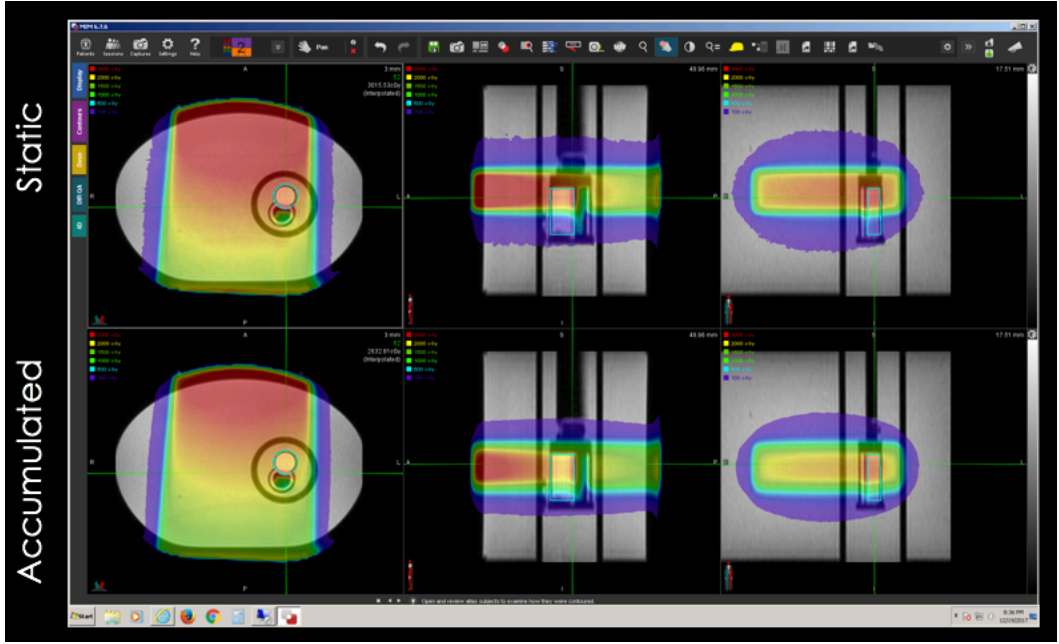

Figure 7. Comparison of 3D dose distributions calculated with incorporation of transverse 1.5T magnetic field. Dose from open 4cm x 15cm field calculated on static target from one 4D-MRI bin and epoch (top row). Accumulated dose from 4cm x 15cm field calculated on each 4D-MRI bin and epoch with MU scaled according to Table 1, registered locally over the target, and summed (bottom row). As expected, dose smearing due to motion is apparent in the accumulated dose.

Figure 8 displays the relationship between spin-lattice relaxation rate, $R_1$, $(1/T_1)$ and absolute dose for the FXG gel. The $R_1$ vs absorbed dose line of best fit is given in Equation 6. Here, the dose in cGy is calculated using the longitudinal relaxation rate in units of $sec^{-1}$. The $R^2$ value of linear fit was 0.995.

$$R_1 = 0.033(Dose) + 0.908 \qquad\qquad [6]$$

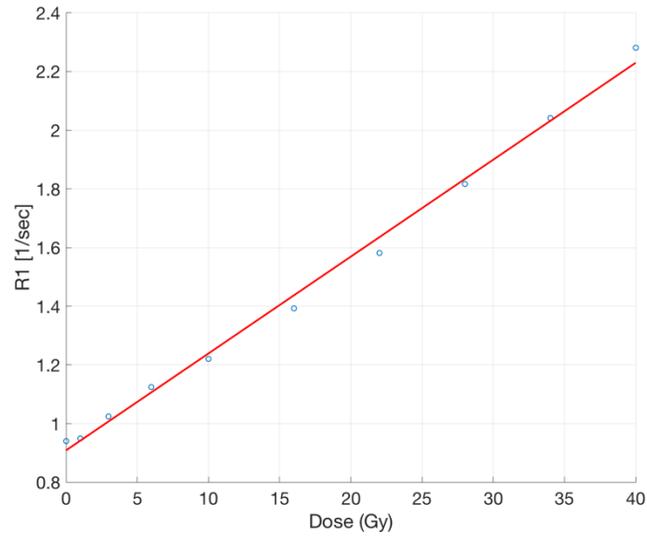

Figure 8. Calibration curve relating absolute dose of FXG gel to longitudinal relaxation rate (R1) demonstrates linear relationship.

Following the calibration experiment, the relationship between absorbed dose and $R_1$ was applied to the 3D $R_1$ map acquired on the irradiated phantom. The 3D dose distribution measured via the FXG gel and the reconstructed dose calculated in MIM are shown in Figure 9. The mean difference between reconstructed and measured dose inside the pancreas and duodenum phantom was 4.4% of the maximum prescribed dose. The localized dose increases at the air-duodenum interface arising from the electron return effect are apparent in both the reconstructed and measured dose distributions. In addition, similar smearing of the dose distributions along the superior/inferior direction due to motion is apparent.

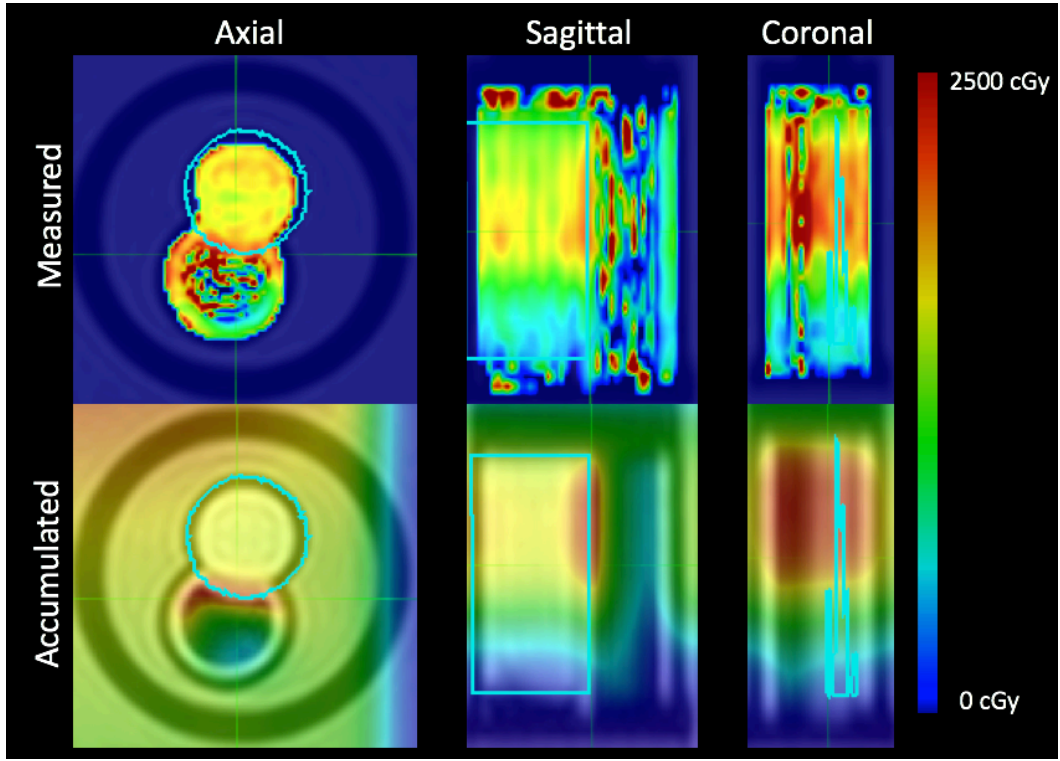

Figure 9. Measured 3D dose distributions obtained via $R_1$ mapping of FXG gel in treated 3D-printed target of dynamic motion phantom (top row) demonstrates good agreement to reconstructed dose (bottom row). Focal dose hot spots due to the electron return effect are observable at the pancreas-duodenum interface.

An example *in vivo* sagittal navigator image of a healthy volunteer is shown in Figure 10a. Due to the high levels of acceleration, the image has a relatively low signal-to-noise ratio. However, sufficient contrast is present to visualize the liver and kidneys, and to extract a respiratory waveform to be used for data sorting.

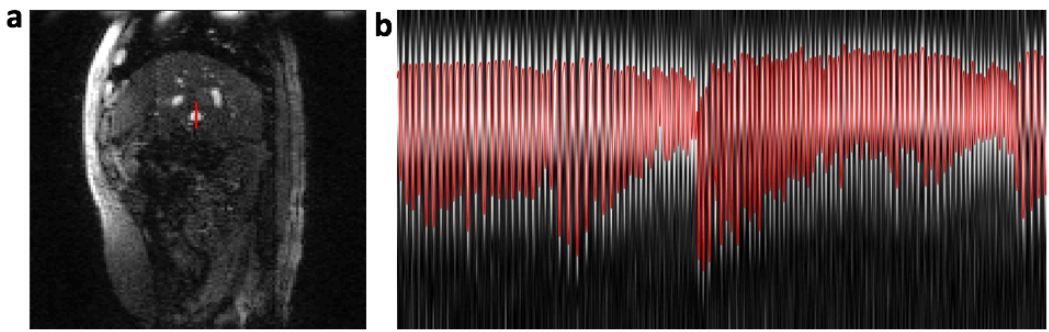

Figure 10. Results from *in vivo* study in a healthy volunteer. Example sagittal navigator image (a) and respiratory motion waveform (red time series in b) extracted from an intensity projection (red line in a) placed in the liver.

A sagittal reformat of end-inspiratory phase images of each of the four 4D-MRI epochs for the 1PL-4D-SOPI volunteer scan are shown in Figure 11. In this volunteer, the peak position of the liver at the end of inspiration remained relatively constant throughout all epochs, as demarcated by the red reference line. Visually, there is enough contrast in the images to delineate the liver and right kidney. In Figure 12, end-inspiratory and end-expiration phase images from the first 4D-MRI epoch are shown.

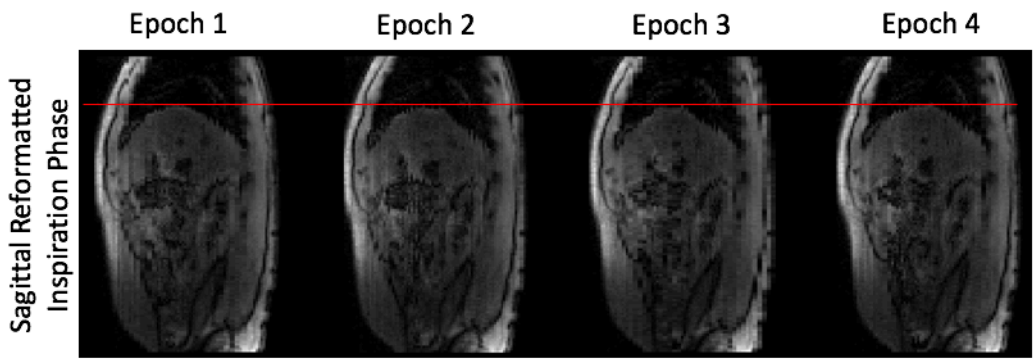

Figure 11. Sagittal reformats of the end-of-inspiration phase for each of the four 4D-MRI epochs from the 1PL-4D-SOPI volunteer scan.

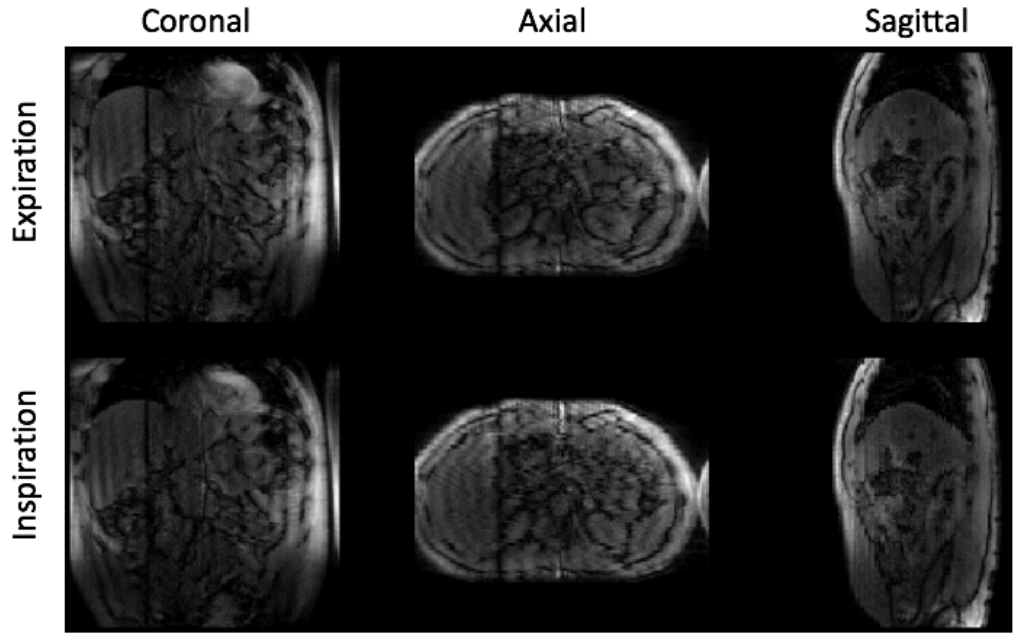

Figure 12. In vivo expiratory (top) and inspiratory (bottom) phase bin reconstructions from one 4D-MRI epoch in a healthy volunteer.

## Discussion

We introduced here a new imaging framework for intrafraction motion management in MR-gRT based on the simultaneous orthogonal plane imaging (SOPI) pulse sequence. The primary advantage of 4D-SOPI for MR-gRT is that it supports simultaneous real-time cine imaging for intrafraction motion management (e.g., breath hold, gating, tracking, trailing) plus construction of serial, self-navigated, respiratory-correlated 4D volumes without DIR for truth-in-delivery analysis.

With 4D-SOPI, DIR algorithms are not required to construct dynamic volumetric images for dose calculation. The construction of serial 4D volumes throughout a treatment fraction permits the 4D-SOPI method to resolve intrafraction motions arising from respiration, organ filling, drifting (e.g., due to muscle relaxation), and bulk motion. Although DIR algorithms are not required for serial 4D-MRI construction with 4D-SOPI, accumulation of dose for truth-in-delivery analysis will still require DIR.

Differences in frequency and directionality challenge the simultaneous resolution of respiratory and peristaltic motions in the abdomen. 4D-SOPI is capable of handling these motions simultaneously through prescription of a coronal navigator slice, which permits self-navigated 4D-MRI reconstruction for dose reconstruction, and enables real-time intrafraction motion management strategies (e.g. gating, tracking) to control for peristaltic motions.

4D-SOPI permits several permutations of navigator and imaging slice geometries. One or more navigator slices can be acquired individually, orthogonally, or in parallel depending on the application. For example, parallel navigator slices could be positioned over the diaphragm and tumor or OAR. The diaphragm slice could be used for image-based, self-navigated 4D-MRI reconstruction while the tumor/OAR slice could be used to gate the radiation delivery. Alternatively, orthogonal navigator slices could be positioned to intersect a tumor or OAR. Similarly, one or more imaging slices can be acquired individually, orthogonally, or in parallel. This unique flexibility also offers the potential to utilize super-resolution 4D-MRI reconstructions [32], [33] with 4D-SOPI, which is an avenue of future exploration.

4D-SOPI offers some flexibility in the contrast of the acquired images. Contrast from any gradient spoiled pulse sequence (e.g. FLASH, FISP, PSIF) can be obtained depending on the desired contrast of the target. Furthermore, when employing the non-equal echo time (nETE) variant of SOPI, the effective TE of the navigator and imaging slice groups can be swapped by changing the order of excitation, thus providing additional control over contrast. This flexibility in image contrast can be exploited to improve the accuracy of DIR for use in dose reconstruction or contour propagation in motion management strategies.

Validation of 4D-MRI has long been a challenge. One validation approach compares motion estimates from 4D-MRI against separate cine MRI acquisitions [16], [18]. A unique advantage of 4D-SOPI for 4D-MRI is self-validation. Average motion contained within the navigator slice can be used to validate motion of the 4D-MRI epoch obtained over the same time period, in the same acquisition.

As seen by slice mismatch in both the phantom and *in vivo* images, small variations in the motion state exist within each 4D-MRI bin. Using more bins would reduce this effect at the expense of requiring the interpolation of more slices that were not filled in each bin. As currently implemented, the image sorting takes place after obtaining magnitude coil-combined images. A potential improvement in image quality could be obtained by performing complex averaging of slices in the same location shuffled into the same bins.

There are a few limitations of this study. First, quantitative comparison of accumulated and measured 3D dose distributions by Gamma analysis was challenged due to Gibbs ringing in the 3D $R_1$ maps (see Fig 9). Though collection of higher resolution DESPOT1 images would rectify this issue, lower resolution $R_1$ mapping was performed to image faster, minimize fading of the FXG gel during measurement. Second, a simple beam model was chosen to validate the use of the 4D-SOPI for dose reconstruction. The relative length of the treatment fraction spent in each of the phases within each of the 4D-MRI epochs was used to weight the dose calculated on each respective 3D volume. For more complicated treatments (e.g., rotational delivery or step-and-shot IMRT), this simple linear weighting will not be accurate. Third, the "intrafraction" motion of the phantom was calculated on a separate MRI scanner from the MR-Linac. Future work will implement the proposed imaging methodology on the Elekta MR-Linac, which will eliminate setup uncertainties associated with moving the motion phantom between systems and timing delays between the start of the motion and the start of the imaging/irradiation.

The 4D-SOPI method draws some similarity to a recently introduced 4D-MRI method that utilizes parallel imaging with controlled aliasing (CAIPIRINHA) to separate simultaneously acquired data from a sagittal navigator slice and a sagittal imaging slice [11]. However, through the use of SOPI and multiband pulses, 4D-SOPI provides more flexibility in navigator and imaging slice prescriptions.

It may be tempting to consider using the 4D-SOPI method to construct motion models for use in treatment planning (i.e., during MR simulation). While this is a possibility, higher quality 4D-MR images may be obtained with dedicated 3D approaches in shorter scan durations [17], [34]. The main benefit of 4D-SOPI is the simultaneous real-time imaging and construction of serial, self-navigated 4D-MR epochs for dose reconstruction for abdominal and thoracic MR-gRT.